\documentclass[10pt, aps, prb, twocolumn, showpacs, showkeys, floatfix]{revtex4-1}
\usepackage{amsmath, amssymb, amsxtra, graphicx, array, mathtools, empheq, color, bbm}
\usepackage{hyperref}
\hypersetup{colorlinks=true, linkcolor=blue, citecolor=blue}
\usepackage{times}

\newcommand{ \ex }				[1] { \langle{#1}\rangle }

\newcommand{\trm}				{\textrm}

\newcommand{\pd}				{\phantom{\dag}}

\newcommand{ \Hc }				{{\rm{H.c.}}}

\newcommand{\nn}				{\nonumber}

\begin{document}


\title{Classical Dynamical Gauge Fields in Optomechanics}

\author{Stefan~Walter $^{1,2}$}
\author{Florian~Marquardt $^{1,2}$}
\affiliation{
$^{1}$ Institute for Theoretical Physics, University Erlangen-N{\"u}rnberg, Staudtstra{\ss}e 7, 91058 Erlangen, Germany \\
$^{2}$ Max Planck Institute for the Science of Light, Staudtstra{\ss}e 2, 91058 Erlangen, Germany
}

\date{\today}


\begin{abstract}
Artificial gauge fields for neutral particles such as photons, recently attracted a lot of attention in various fields ranging from
photonic crystals to ultracold atoms in optical lattices to optomechanical arrays. Here we point out that, among all implementations
of gauge fields, the optomechanical setting allows for the most natural extension where the gauge field becomes dynamical.
The mechanical oscillation phases determine the effective artificial magnetic field for the photons, and once these phases
are allowed to evolve, they respond to the flow of photons in the structure. We discuss a simple three-site model where we
identify four different regimes of the gauge-field dynamics. Furthermore, we extend the discussion to a two-dimensional
lattice. Our proposed scheme could for instance be implemented using optomechanical crystals.
\end{abstract}

\maketitle

\section{Introduction}
Optomechanics describes the interaction of light and mechanical motion~\cite{Aspelmeyer2014}. The prototypical
optomechanical setting consists in a Fabry-P{\'e}rot cavity where one of the mirrors is free to oscillate. Due to the
radiation pressure force the light inside the cavity interacts with the mirror's motion. 
Tremendous experimental progress has been made during the last years to exploit this very elementary light-matter
interaction, with achievements such as cooling a nanomechanical oscillator to its motional ground state~\cite{Teufel2011,Chan2011a}
and position measurements below the standard quantum limit~\cite{Teufel2009}, to name only a few examples.
Mechanically and/or optically coupling several optomechanical systems leads to interesting new physics. For
instance, setups consisting of only a few optical and mechanical modes allow for nonreciprocal devices for
photons~\cite{Manipatruni2009, Habraken2012, Hafezi2012, Wang2015, Ruesink2015, Fang2016}. Furthermore, one- or two-dimensional arrays of
coupled optomechanical systems are promising candidate systems for studying many-body physics of photons
or phonons~\cite{Bhattacharya2008,Chang2011,Xuereb2012, Tomadin2012,Schmidt2012,Akram2012,Chen2012,Schmidt2015a,Schmidt2015b,Peano2014}.
Most interestingly, optomechanical arrays are also a platform to create artificial gauge fields for photons~\cite{Schmidt2015b}
and phonons~\cite{Peano2014}. The optomechanical implementation complements other proposals for generating artificial
gauge fields for photons~\cite{Haldane2008, Wang2009, Koch2010, Umucallar2011, Hafezi2011, Fang2012b, Hafezi2013, Rechtsman2013, Lu2014, Mittal2014}
and ultracold atoms in optical lattices~\cite{Jaksch2003, Sorensen2005, Lin2009, Aidelsburger2011, Aidelsburger2013, Jotzu2014}.

In this article, we study the most basic phonon-assisted photon tunneling process which is due to the optomechanical
interaction. We show that an elementary optomechanical setting naturally gives rise to \emph{dynamical} gauge
fields. The key ingredient is a self-oscillating mechanical mode which connects two optical modes. Most importantly
this brings an additional degree of freedom into play, viz. the mechanical oscillation phase. This mechanical
oscillation phase is connected to an effective magnetic field seen by the photons and possesses its own dynamics.

During the last years, several proposal have been put forward dealing with the deliberate generation of dynamical
gauge fields. Platforms based on ultracold atoms in optical lattices~\cite{Osterloh2005,Ruseckas2005,Zohar2011,Hauke2012,Banerjee2012,Zohar2013,Edmonds2013}
or superconducting circuits~\cite{Marcos2013,Marcos2014} are discussed as promising systems which could serve as
quantum simulators for dynamical gauge theories such as quantum electrodynamics and quantum chromodynamics~\cite{Wiese2013,Zohar2015}.
In most cases these proposals require a great deal of engineering, meaning carefully choosing a setting which yields
the desired interactions between for instance superconducting qubits. Instead we concentrate on the intriguing optomechanical
setting which gives rise to dynamical gauge fields in a very natural way. Particularly, in our scenario only basic phonon-assisted
photon tunneling processes generated via the basic optomechanical interaction are needed.
Besides the purpose of quantum simulation for dynamical gauge theories, the optomechanical setting opens up new
directions dealing with nonlinear pattern formation of dynamical gauge fields in driven and dissipative systems~\cite{Lauter2015}. 

In the following, we investigate the evolution of the mechanical oscillation phases (the dynamical gauge field) in response
to the light field dynamics, which is a unique feature of the optomechanical case.
We consider a photonic lattice (representing the matter fields) and artificial gauge fields (phonons) which can be
attributed to directed links between two sites of the photonic lattice. Such a system could for instance be implemented in
optomechanical crystal structures~\cite{Eichenfield2009,Safavi2010a,Safavi2010b,Gavartin2011,Chan2011a,Safavi2014}
or disk resonator arrays~\cite{Zhang2015}.

\section{Generic Model for Dynamical Gauge Fields with Optomechanics}
The crucial ingredient in our model is the phonon-assisted photon tunneling that can be generated by the optomechanical
interaction: A photon $a$ hopping from site to site is accompanied by the coherent emission or absorption of a phonon $b$.
This can be described by the Hamiltonian ($\hbar=1$)
\begin{align}\label{eqn:eq1}
	H =	\sum_{j} \nu_{j}^{\pd} a_{j}^{\dag} a_{j}^{\pd} + \sum_{l} \omega_{l}^{\pd} b_{l}^{\dag} b_{l}^{\pd} + \sum_{l} J_{l}^{\pd} b_{l}^{\pd} a_{l_{2}}^{\dag} a_{l_{1}}^{\pd} + \Hc \, .
\end{align}
Here, $j$ denotes a lattice site, and $l=(l_{1},l_{2})$ is the index for a \emph{directed} link from $l_{1}$ to $l_{2}$.
A photon hopping in the direction of the link absorbs a phonon. Photons (phonons) have frequencies $\nu_{j}$
($\omega_{l}$), and $J_{l}$ are the phonon-assisted photon tunneling amplitudes. We introduce the notation
$b_{ij} = b_{ji}^{\dag}$ and note that we use $b_{l}$ and $b_{l_{1} l_{2}}$ interchangeably. In Eq.~(\ref{eqn:eq1})
we made use of the rotating wave approximation which is valid for $\omega_{l} >\kappa, J_{l}$, where $\kappa$ is the photon decay rate.  
Non-reciprocity in photon transport can be engineered by coherent inelastic transitions induced by mechanical
vibrations~\cite{Schmidt2015b}. However, in contrast to Ref.~\onlinecite{Schmidt2015b} we will treat the vibrations as
dynamical degrees of freedom. In order to make the inelastic processes resonant, the nearest neighbor on-site
photon frequencies have to differ by the corresponding link phonon frequency ($\omega_{ij} = \nu_{j} - \nu_{i}$)
with the link direction from site $i$ to $j$. This leads to \emph{directed} links. Generally speaking, a photon tunneling
from a site with a low (high) on-site frequency to a site with a high (low) on-site frequency absorbs (emits) a phonon.
We include photon losses at a rate $\kappa$ and account for driving by adding the term
$H_{d} = \sum_{j} E_{j} \left( a_{j} e^{i \omega_{d} t}+ a_{j}^{\dag} e^{-i \omega_{d} t} \right)$ to Eq.~(\ref{eqn:eq1}),
see~\ref{appB}.

%
\begin{figure}[ht]
\begin{center}
	\includegraphics[width=0.95\columnwidth]{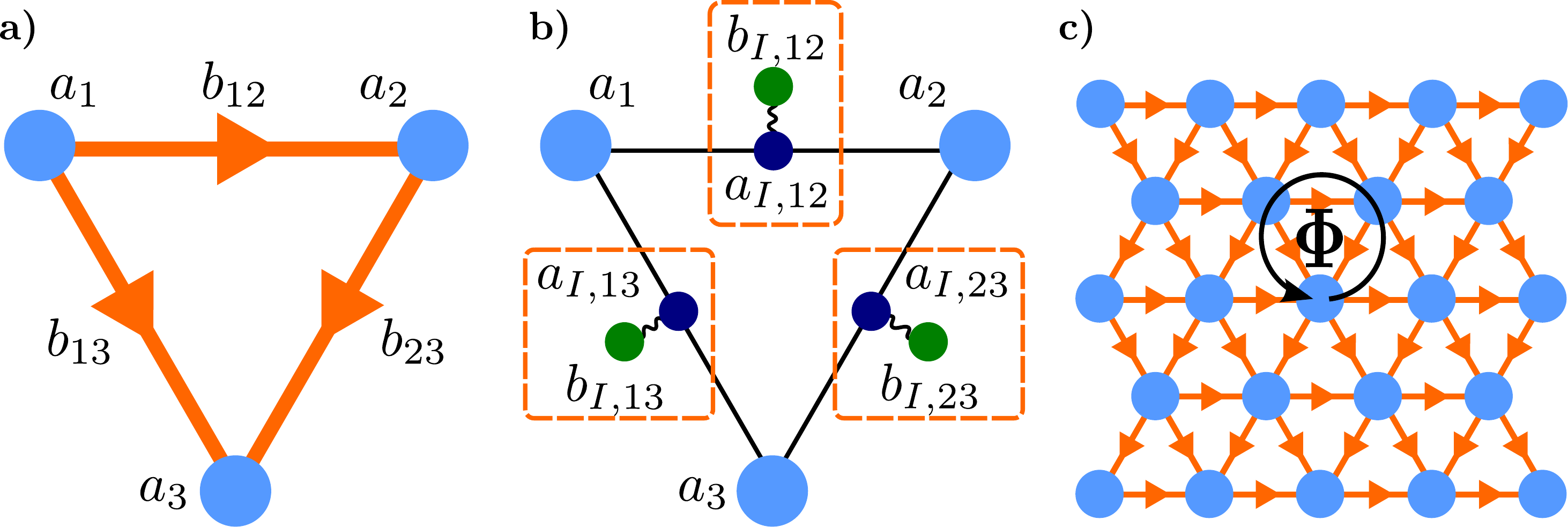}
	\caption{
	a) Effective model for a three-site optomechanical implementation of dynamical gauge fields for photons $a$ (light blue)
	coupling to phonons $b$ (orange). For the optical frequencies we choose $\nu_{1}<\nu_{2}<\nu_{3}$ leading to
	directed links, i.e., photons hopping in direction of the orange arrow absorb a phonon.
	b) Possible realization making use of the so-called ``modulated link'' scheme. Dark blue (green) dots depict intermediate
	optical (mechanical) modes which couple optomechanically to form an optomechanical cell (dashed box).
	c) Two-dimensional triangular lattice.
	}
	\label{fig:om_schematics}
\end{center}
\end{figure}

Our effective model,~Fig.~\ref{fig:om_schematics} a), can be realized in an optomechanical setting by building on the
``modulated link'' scheme that had been proposed to generate \emph{static} artificial magnetic fields~\cite{Schmidt2015b}.
In this scheme, the link between two optical modes $i$ and $j$ is realized with an intermediate optical mode $a_{I,l}$ which
couples optomechanically to a mechanical mode $b_{I,l}$, forming a single optomechanical cell, cf. Fig.~\ref{fig:om_schematics} b).
In Ref.~\onlinecite{Schmidt2015b} the mechanical mode is externally driven into a large amplitude state $\ex{b_{I,l}(t)} = B_{I,l} e^{-i (\Omega_{I,l} t +\phi)}$
which leads to a modulation of the frequency of the intermediate optical mode $\omega_{I,j}(t) = \omega_{I,j} +2 g_{0} B_{I,l} \cos(\Omega_{I,l}t + \phi)$.
$g_{0}$ is the single-photon optomechanical coupling strength. We note that other microscopic implementations are possible.
For example, one might have a mechanical mode that directly couples to the hopping between optical modes, as has been
worked out in detail for optomechanical crystals~\cite{Safavi2011}. This would be connected to the ``wavelength conversion
scheme'' discussed in Ref.~\onlinecite{Schmidt2015b}.

In contrast to these two scenarios, we will assume that the mechanical oscillator undergoes self-sustained optomechanical
oscillations~\cite{Aspelmeyer2014} instead of being externally driven. Thus it behaves as a limit-cycle oscillator with a
fixed amplitude $B$ and a free phase $\phi$. We will later show that the phase of the mechanical limit-cycle oscillator
posses its own dynamics and that it will respond to the flow of photons in the system. This phase will then be directly
linked to a dynamical gauge field. We want to stress that such mechanical self-sustained oscillations are a very generic
feature of optomechnical systems. Therefore an optomechanical implementation of a dynamical gauge field requires
less engineering than other proposals~\cite{Osterloh2005,Ruseckas2005,Zohar2011,Hauke2012,Banerjee2012,Zohar2013,Edmonds2013,Marcos2013,Marcos2014}.

\section{Gauge Field Dynamics}
Here we will focus on the classical dynamics of the model, i.e., the limit of large coherent photon and phonon amplitudes.
This is the most relevant regime for most of the current optomechanical setups (due to the small single-photon coupling
strength $g_{0}$)~\cite{Aspelmeyer2014}. Thus, we decompose the expectation values of the photon and phonon operators
into a classical amplitude and a phase, $a_{j} = A_{j} e^{i\theta_{j}}$ and $b_{ij} = B_{ij} e^{i \phi_{ij}}$. From the full quantum
Heisenberg equations of motion for the effective Hamiltonian, the equations for the mechanical phases $\phi_{ij}$ become
\begin{align}\label{eqn:eqm1}
	\dot{\phi}_{ij}	&= -\omega_{ij} - \frac{J_{ij} }{B_{ij}} A_{i} A_{j} \cos(\phi_{ij}+\theta_{ij}) \, ,
\end{align}
while the optical amplitudes $A_{i}$ obey
\begin{align}\label{eqn:eqm2}
	\dot{A}_{i} =  -\sum_{j \neq i} J_{ij}  B_{ij} \sin(\phi_{ij} + \theta_{ij}) A_{j} \, ,
\end{align}
and the optical phases $\theta_{i}$ evolve according to
\begin{align}\label{eqn:eqm3}
	\dot{\theta}_{i}	= - \nu_{i} - \sum_{j \neq i} J_{ij}  B_{ij} \frac{A_{j}}{A_{i}} \cos(\phi_{ij} + \theta_{ij}) \, .
\end{align}
We introduced $\theta_{jk} = \theta_{j} - \theta_{k}$ and used $\phi_{ij} = -\phi_{ji}$.
Here and in the following we assume that the amplitudes of the limit-cycle oscillations are a constant of motion,
i.e., $\dot{B}_{ij}=\dot{B}_{ji}=0$. This regime can be reached by working with self-induced optomechanical oscillators sufficiently
above threshold~\cite{Marquardt2006}. The initial phase of the self-oscillators would be random
without extra precautions, but it can be set via an externally imposed mechanical drive, realized through an intensity-modulated
light field.

In principle, the quantum regime of the present model could also be discussed, if the optomechanical coupling would be strong.
The most straightforward extension to a quantum version can be realized considering that the mechanical oscillators on the
links feature limit-cycles with quantum coherent phase dynamics.
Another more demanding way towards dynamical gauge fields in the quantum regime with optomechanics would be in the
spirit of Refs.~\onlinecite{Osterloh2005,Ruseckas2005,Zohar2011,Hauke2012,Banerjee2012,Zohar2013,Edmonds2013,Marcos2013,Marcos2014}.
There, the link variable can be considered as a spin that flips each time an excitation hops between the corresponding sites.
In the optomechanical case, this would require highly nonlinear mechanical oscillators that can be brought into the quantum
regime such that they effectively act as a two-level system.

An important point we want to make at this stage is the invariance of the equations of motion under the following
\emph{local} $U(1)$ gauge transformation
\begin{align}
	\phi_{ij}^{\prime}	&= \phi_{ij} + \left( \chi_{j} - \chi_{i} \right) \label{eqn:g1} \, , \\
	\theta_{i}^{\prime}	&= \theta_{i} + \chi_{i}  \label{eqn:g2} \, ,
\end{align}
meaning that the observed evolution of the light intensity will not change, independent of the choice of $\chi_{j}$.
We assumed a static gauge choice $\chi$; otherwise Eqs.~(\ref{eqn:g1}) and~(\ref{eqn:g2}) would have to be supplemented by a change in
frequencies: $\omega_{ij}' = \omega_{ij} + \dot{\chi}_{i} - \dot{\chi}_{j}$ and $\nu_{i}' = \nu_{i} - \dot{\chi}_{i}$.

Under which conditions does the gauge field display nontrivial dynamics? In Eqs.~(\ref{eqn:eqm1}), ~(\ref{eqn:eqm2}),
and~(\ref{eqn:eqm3}) we can rescale time by $J B$, where we assume link independent tunneling and mechanical amplitudes.
Doing this, we observe that the entire dynamics only depends on the dimensionless ratio of $A/B$. Here, $A$ is proportional to
the laser drive amplitude and corresponds to the optical amplitude at an arbitrary reference site in the lattice (globally lowering
the optical amplitudes will also lower the optical amplitude at the reference site). For the limit $A/B \to 0$, we expect that the
oscillation phases will not be affected by the hopping photons, rather only defining a static magnetic field pattern. This can
be seen from Eq.~(\ref{eqn:eqm1}): for $A/B \ll 1$ the second term, which provides the coupling to the hopping photons, can
be neglected. However, if $A/B$ is large, we expect back-action of the hopping photons on the phonons leading to
intriguing coupled dynamics of the gauge field.

\section{Three-Site Model}

First, we study the case of three sites. The resulting effective model for photons $a_{j}$ on sites $j \in \{1,2,3\}$ and
phonons $b_{l}$ on links $l \in \{12,23,13\}$ is depicted in Fig.~\ref{fig:om_schematics} a), where we define $a_{4} = a_{1}$.
For definiteness, we will assume $\nu_{1} < \nu_{2} < \nu_{3}$. For three sites, the gauge freedom implies that only the gauge
invariant flux
\begin{align}
	\Phi = \phi_{13} + \phi_{21} + \phi_{32} \, ,
\end{align}
i.e., the sum of phases around the triangular plaquette, affects the dynamics of the photons. We want to mention that in
the regime where the phonons are not influenced by the photons ($\Phi=\trm{const.}$) the Hamiltonian for the three-site
model can be diagonalized and the setup can act (for $\Phi = \pm \pi/2$) as a photon circulator~\cite{Koch2010}, see also~\ref{appA}.
%

\begin{figure}[ht]
\begin{center}
	\includegraphics[width=0.95\columnwidth]{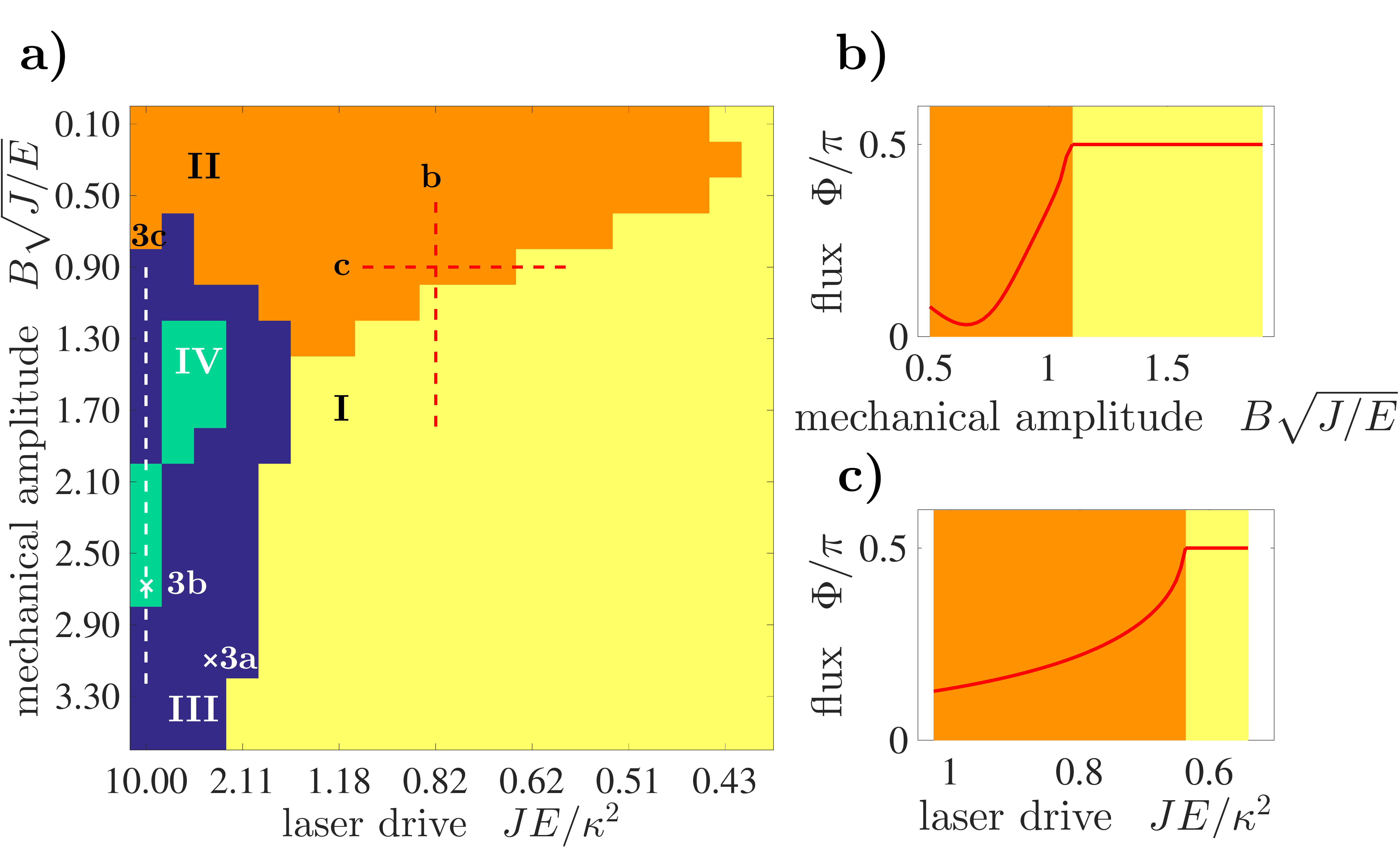}
	\caption{
	a) Dynamical regimes of the flux $\Phi$. Phase diagram as a function of mechanical amplitude and laser drive
	(which is resonant on site $1$, $\omega_{d} = \nu_{1}$) showing four regimes for the flux dynamics.
	In regimes I and II the flux $\Phi$ is stationary and tends to a value equal to $\pi/2$ or different from it, respectively.
	In regimes III and IV the flux $\Phi$ is dynamical and can either show a periodic oscillatory or chaotic behavior, respectively.
	b) and c) Cuts along the red dashed lines in a) across the phase transition from regime I to II.
	}
	\label{fig:fig21}
\end{center}
\end{figure}
%
%
\begin{figure}[ht]
\begin{center}
	\includegraphics[width=0.8\columnwidth]{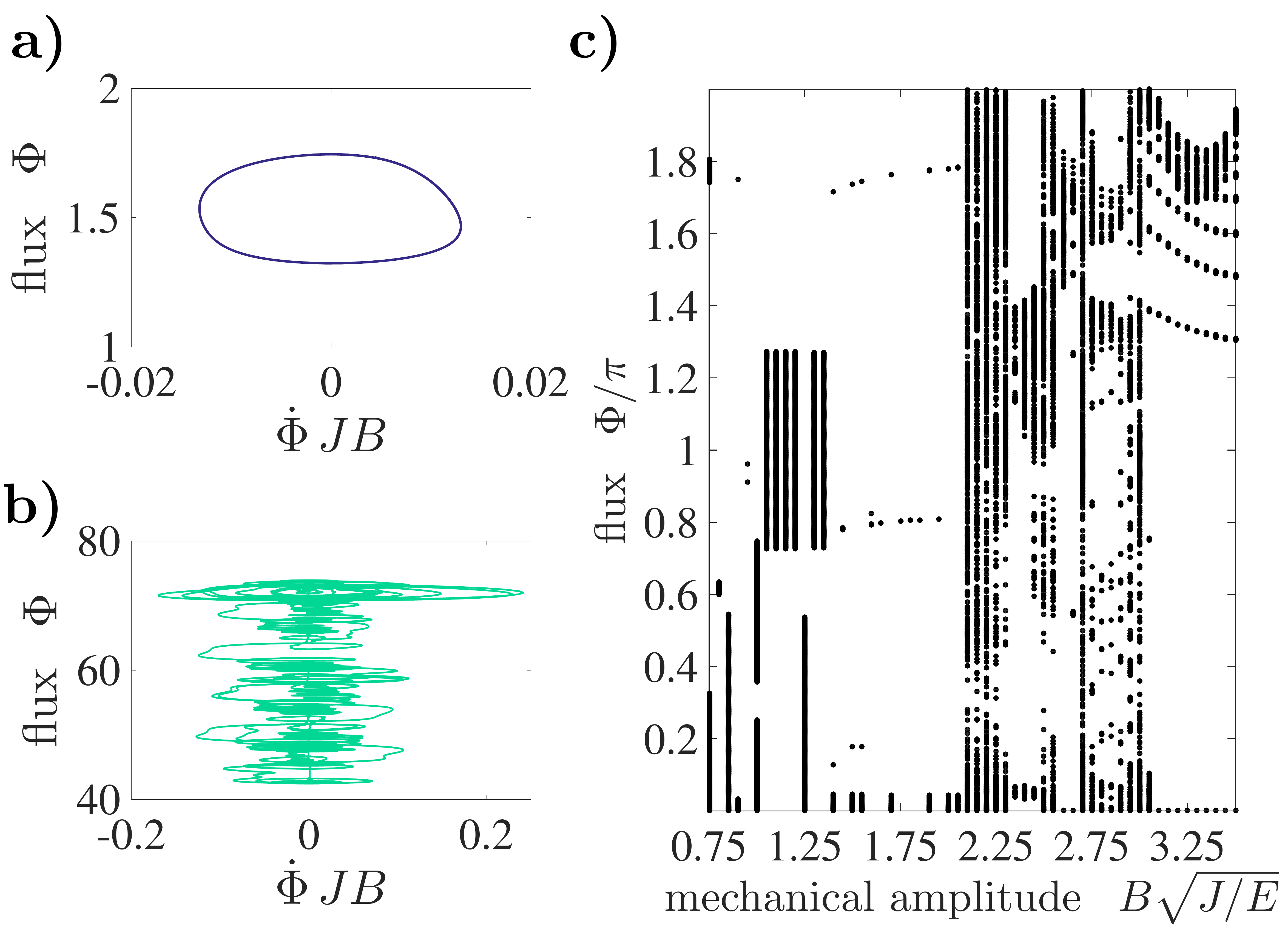}
	\caption{
	Examples of phase space trajectories $\Phi(t)-\dot{\Phi}(t)$ in region III and IV are shown in a) and b), respectively.
	c) Bifurcation diagram for the flux dynamics. We display the values of $\Phi(t)$ attained at the zero-crossings of $\dot{\Phi}(t)$.
	For the bifurcation diagram c) we used a higher resolution for values of $B \sqrt{J/E}$ than for the phase diagram in Fig.~\ref{fig:fig21} a).
	}
	\label{fig:fig22}
\end{center}
\end{figure}
In the intriguing case where the photons interact with the phonons we discuss the driven and dissipative setting and
furthermore choose equal tunneling amplitudes $J_{l}=J$ and mechanical amplitudes $B_{l} =B$.
As an example we consider a resonant drive on site $j=1$.
Without drive and dissipation the dynamics only depends on the ratio $A/B$. As a consequence,
for the driven case, the four parameters involved $(E, \kappa, B, J)$, can be combined into just two dimensionless parameters,
$B \sqrt{J/E}$ and $J E/\kappa^{2}$. The resulting ``phase diagram'' for the flux dynamics as a function of these two parameters
is displayed in Fig.~\ref{fig:fig21} a). It has been obtained from direct numerical simulations and reveals four distinct regimes.
In regimes I and II the flux $\Phi(t)$ (after some transient behavior) approaches a stationary value of either $\Phi(t \to \infty) = \pi/2$
or different from it, respectively. In regimes III and IV, the flux $\Phi(t)$ is not stationary but even in the long-time limit
shows a dynamical behavior. Most interestingly, the flux dynamics $\Phi(t)$ in these regimes can either show periodic
or chaotic behavior, region III and IV in Fig.~\ref{fig:fig21} a), respectively. Figures~\ref{fig:fig21} b) and c) show cuts along
the red dashed lines marked in the phase diagram, indicating a continuous phase transition from phase I to II. In Fig.~\ref{fig:fig22}
a) and b), we show two examples of the phase space $\Phi(t) - \dot{\Phi}(t)$ in regimes III and IV. Already at this level
we can distinguish periodic [Fig.~\ref{fig:fig22} a)] from chaotic [Fig.~\ref{fig:fig22} b)] dynamics. A more involved characterization
can be done using a bifurcation diagram. To this end, we show the value of $\Phi(t)$ evaluated at the zero crossings of
$\dot{\Phi}(t)$, in the long-time limit, as a function the mechanical amplitude $B$ in Fig.~\ref{fig:fig22} c). This bifurcation
diagram allows us to distinguish the periodic from the chaotic flux dynamics within the whole phase diagram for $\Phi(t)$.
In addition, we also checked whether the Fourier transform of the trajectories shows a clear peak or is flat, indicating
periodic or chaotic behavior, respectively, see~\ref{appC}.

In the regime of fast photons dynamics (compared to the phonon dynamics), we are able to apply a Born-Oppenheimer
approximation and adiabatically eliminate the photons. To be more precise, we solve $da_{i} /dt =0$ where
$\dot{a}_{j} = \left[-i (\nu_{j} - \omega_{d}) - \frac{\kappa}{2}\right] a_{j} - i E_{j} - i J B \sum_{k \neq j} e^{-i\phi_{jk}} a_{k}$
and use this instantaneous solution to eliminate $a_i$ from the equations of motion for $\phi_{ij}$, see~\ref{appA}.
In the case of a resonant drive on site $j=1$, this approximation leads to the following equation of motion for the flux:
\begin{align}
	\dot{\Phi} = \frac{16 E^{2}}{J B^{3}} \frac{(4 + \frac{\kappa^{2}}{J^{2} B^{2}}) \cos(\Phi)} {\frac{\kappa^{2}}{J^{2}B^{2}} [12 + \frac{\kappa^{2}}{J^{2} B^{2}}]^{2} + [16 \cos(\Phi)]^{2}} \label{eqn:dP} \, .
\end{align}
From Eq.~(\ref{eqn:dP}) we find $\Phi(t \to \infty) = \pi/2$ which shows very good agreement with the exact numerical
long-time dynamics in regime I. This approach fails in the other regimes since there we are not able to adiabatically
eliminate the photons. We also want to mention that both $\Phi = + \pi/2$ and $\Phi = - \pi/2$ are fixed points of Eq.~(\ref{eqn:dP}).
It turns out that for a resonant drive on site $j=1$, $\Phi = - \pi/2$ is an unstable fixed point of Eq.~(\ref{eqn:dP}).
The asymmetry between $+\pi/2$ and $-\pi/2$ is due to the breaking of translational invariance, necessarily produced
by the link directions. In contrast, a resonant drive on site $j=2 \, (3)$ would have $\Phi =- \pi/2 \, (+\pi/2)$ as a stable
fixed point. From Eq.~(\ref{eqn:dP}) we also can estimate the rate $\Gamma$ at which the flux $\Phi(t)$ settles into
steady state. By linearizing around the fixed point we find $\Gamma = [16 E^{2} (4 + \xi) ] / [J B^{3} \xi (12 + \xi)^2)]$
where $\xi = (\kappa/J B)^{2}$.

\section{Lattices}
\begin{figure}[ht!]
\begin{center}
	\includegraphics[width=0.95\columnwidth]{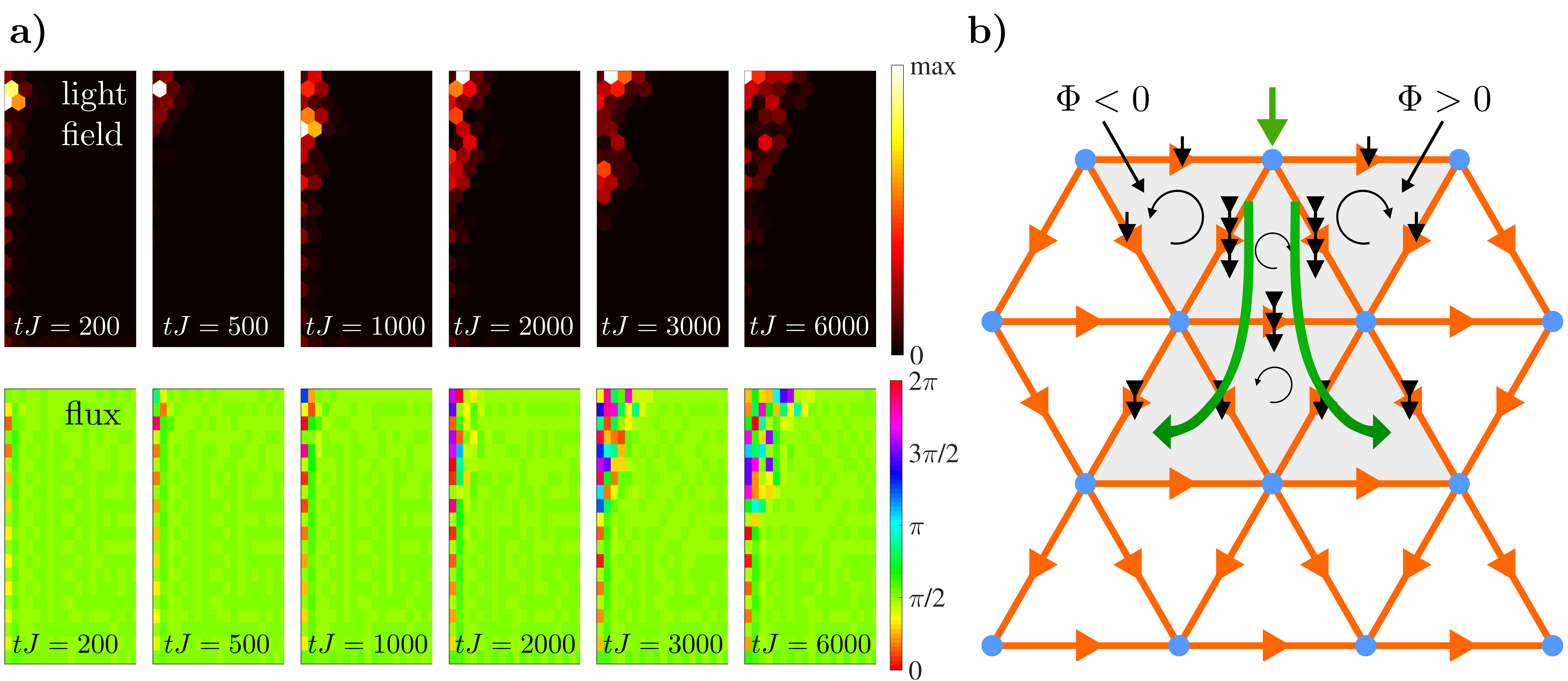}
	\caption{
	a) Evolution of the light field on a lattice with a dynamical gauge field of vibrational origin. Light intensity $A_{i,j}^{2}$
	on each site (top row) and flux $\Phi$ per plaquette (bottom row) at different times. The resonant drive
	is on the site on the corner and the initial flux per plaquette is $\Phi=\pi/2$. No steady state of either photons or
	phonons is reached. We chose $B \sqrt{J/E}= 3.16$, $J E/\kappa^{2}=10$ and have $1/\kappa = 1/J$.
	The simulations were carried out on a $21 \times 21$ triangular lattice, where only the left half is shown.
	b) Schematics of the flux and light field dynamics for the very first moments of the time evolution. Initially $\Phi=0$
	for every plaquette and a single optical site is driven (green arrow).
	A finite optical amplitude on neighboring sites changes the mechanical phases, indicated by the black arrows.
	The number of arrowheads indicates the magnitude of decrease. The link directions lead to a negative (positive)
	flux through the left (right) plaquette next to the illuminated site. The direction of light transport is indicated
	by the green arrows.
	}
	\label{fig:drivedissdyn}
\end{center}
\end{figure}

We extend the three-site model to two-dimensional lattices and illustrate the dynamical behaviour on a triangular lattice,
cf. Fig.~\ref{fig:om_schematics} c). Going from three sites to a lattice, a significant new feature comes into play: the artificial
dynamical magnetic field produced by the phonons can now exert a Lorentz force that bends the path of any light-beam
propagating in the array. Therefore, we end up with a dynamical interplay where the flow of the photons
changes the spatial distribution of the magnetic flux density which then acts back on the dynamics of the light field.
We choose a scenario with the link directions as depicted in Fig.~\ref{fig:om_schematics} c).
We note that this is not the only possible choice. In fact the intricate photon and phonon dynamics depends on the link pattern.
We illustrate the nonlinear structure formation in this model for the case of having only a single site illuminated by a laser.
In Fig.~\ref{fig:drivedissdyn} a) we show the temporal evolution of the light intensity (top row) as well as the magnetic field
(bottom row) on the lattice.
At first the photons experience a static magnetic field which is set by the initial phases of the mechanical oscillations (here
chosen such that the initial flux is $\Phi = \pi/2$) and start to move along the edge. Due to the back-action of the photons
(which primarily move along the edge) on the phonons, the flux per plaquette changes. This in turn leads to a reconfiguration
of the magnetic field which in this scenario forces the photons to reverse their direction of motion. Here, the system does not
reach a steady state even in the long-time limit.
Even though the photons live only for a short time $1/\kappa$ before escaping the structure, the system
develops a spatial ``memory'' in the form of the mechanical oscillation phases, where previous photons leave their imprint.

%

For an intuitive description of the dynamics we assume zero initial flux per plaquette $\Phi = 0$, a large optical damping
$\kappa$ compared to the tunneling $J$, and a drive on one optical site on the top edge of the lattice. For this scenario,
we show in Fig.~\ref{fig:drivedissdyn} b) a schematics of the flux and light field dynamics and focus on the fluxes through
the gray plaquettes for the very first moments of time evolution.
As soon as the optical amplitudes of two neighboring optical sites is nonzero, the mechanical oscillation phase on the
corresponding link will change, cf. Eq.~(\ref{eqn:eqm1}). More precisely, the phases on the links decrease, indicated by
the black arrows where the number of arrowheads indicate the magnitude of decrease.
However, from the way these phases enter the photon dynamics, one can deduce that each phase contributes to the
flux inside a plaquette with a positive sign only if the respective link is traversed in the positive direction when going
around the plaquette counterclockwise. This leads to the signs shown in the figure.
The light propagation due to the magnetic field is indicated by the green arrows.
It is worth mentioning that even in the long-time limit the mechanical oscillation phases continue drifting because even
an arbitrarily small but finite optical amplitude on neighboring sites is enough to change the phases, albeit very slowly.
With such a scenario one could imagine to engineer a desired magnetic field pattern by means of an optical drive.

\section{Disorder}
In general, disorder effects in optomechanical arrays have been recognized as an important issue, especially in systems
where the coupling between optical modes cannot be larger than the $\trm{GHz}$-scale mechanical frequencies, as is
the case both in the present system as well as, e.g., in our proposal for topologically protected transport in optomechanics~\cite{Peano2014}.
The transport of photons and phonons in disordered optomechanical arrays in the regime of Anderson localization has recently
been analyzed in some detail by our group~\cite{Roque2016}. To avoid such effects, the disorder needs to be suppressed
down to a level of the photon tunneling or less in order to obtain localization lengths that are so large that they do not matter
any more (larger than either the system size or the photon decay length). However, recently significant progress has been
made regarding the fabrication of optomechanical crystal arrays, representing the most promising integrated nanoscale
platform for these types of experiments. Specifically, it has been possible to reduce the disorder by a factor of about $100$
using novel postprocessing techniques, which brings it down to a scale where photon localization lengths become very large.
These techniques have been exploited recently in an experiment on optomechanical non-reciprocity with two coupled modes~\cite{Fang2016}
which is quite close in spirit to the setup we would require here.

\section{Conclusion}
We have shown that dynamical gauge fields in optomechanical arrays arise quite naturally. The evolving mechanical
oscillation phases, which respond to the flow of the photons, represent a dynamical gauge field for the latter. Already the
three-site model shows intriguing dynamics which leads to a rather complex phase diagram for the flux dynamics. With
experiments pushing towards multi-mode optomechanical setups, the three-site model seems feasible to be realized in
the near future and would pave the way for further studies of dynamical gauge fields in optomechanical arrays.
Collective behavior such as synchronization and pattern formation of mechanical limit-cycle oscillators have recently been
studied in optomechanical arrays~\cite{Heinrich2011, Ludwig2013, Lauter2015}. In this spirit, further studies on gauge field
dynamics in optomechanics could address questions on synchronization and dynamical pattern formation of the magnetic
field.

We acknowledge helpful discussions with Roland Lauter, and thank Vittorio Peano and Talitha Weiss for a careful reading
of the manuscript. This work was financially supported by the Marie Curie ITN cQOM and the ERC OPTOMECH.

\onecolumngrid
\appendix

\section{The Three-Site Model: Including drive and dissipation}\label{appB}

\begin{figure}[ht]
\begin{center}
	\includegraphics[width=0.8\columnwidth]{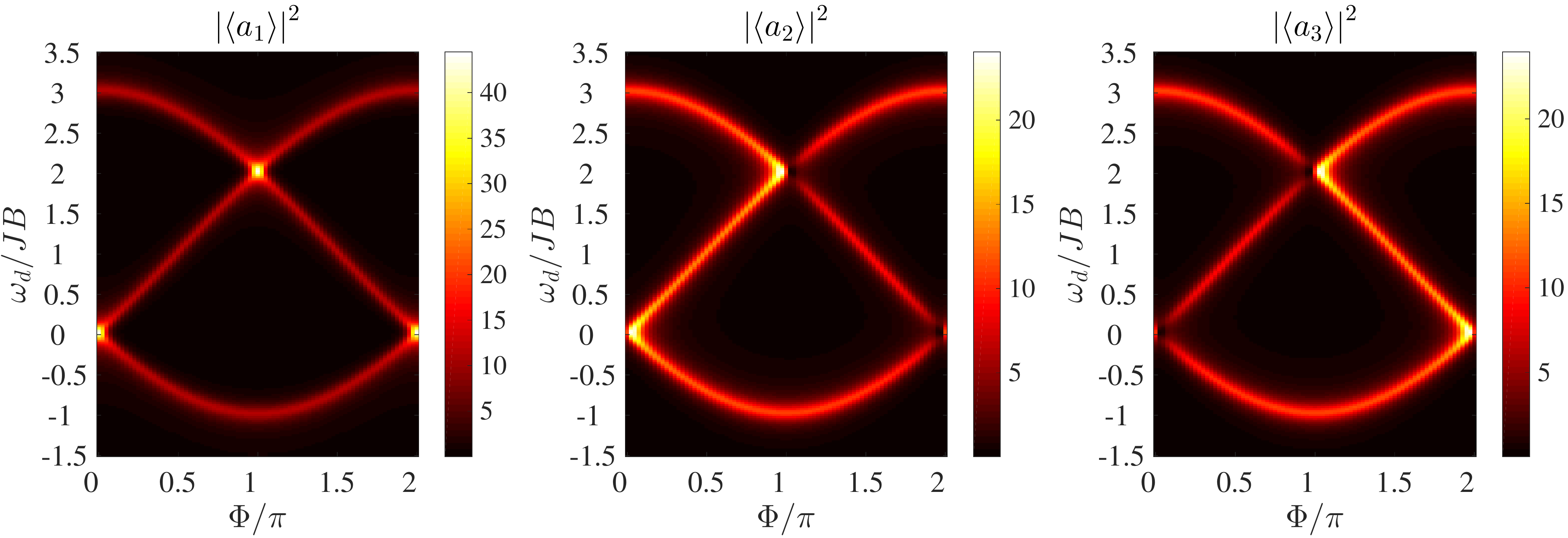}
	\caption{
	Left: $\left| A_{1} \right|^{2}$. Middle: $\left| A_{2} \right|^{2}$. Right: $\left| A_{3} \right|^{2}$.
	Here we only drive site $j=1$, i.e., $E_{1} = E, E_{2}=E_{3}=0$ and choose $J E / \kappa^{2} = 0.025$, and $B\sqrt{J/E} = 31.6$.
	}
	\label{fig:OKdiss}
\end{center}
\end{figure}

Since the photons eventually decay, we add photon loss and drive to the system. This is best done starting with the
Hamiltonian~(\ref{eqn:S1}) and adding a driving term $H_{d} = \sum_{j \in \{1,2,3\}} E_{j} \left( a_{j} e^{i \omega_{d} t}+ a_{j}^{\dag} e^{-i \omega_{d} t} \right)$.
Going into a frame rotating with the driving frequency $\omega_{d}$, we obtain
\begin{align}
	H &=	\sum_{j \in \{1,2,3\}} (\nu_{j}^{\pd} - \omega_{d}) a_{j}^{\dag} a_{j}^{\pd} + \sum_{j \in \{1,2,3\}} E_{j} \left( a_{j} + a_{j}^{\dag} \right) + \sum_{l \in \{12,23,13\}} \omega_{l}^{\pd} b_{l}^{\dag} b_{l}^{\pd} \\
		&+J_{12}  B_{12} \left(  e^{-i \phi_{12}} a_{1}^{\dag} a_{2}^{\pd} + \Hc \right)
		+ J_{23}  B_{23} \left( e^{-i \phi_{23}} a_{2}^{\dag} a_{3}^{\pd} + \Hc \right)
		+ J_{13}  B_{13} \left( e^{i \phi_{13}} a_{3}^{\dag} a_{1}^{\pd} + \Hc \right) \, . \nn
\end{align}
Including dissipation of the photons at a rate $\kappa$ and neglecting effects due to quantum noise, the equations of motion
for the photons become $d\vec{a}/dt = - i M \vec{a} - i \vec{E}$ with
\begin{align}
	\vec{a} &= (a_{1},a_{2},a_{3})^{\trm{T}} \, , \\
	\vec{E} &= (E_{1},E_{2},E_{3})^{\trm{T}} \, , \\
	M& = \left(\begin{array}{ccc}
	(\nu_{1} - \omega_{d}) - i \frac{\kappa}{2} & J_{12} B_{12} e^{-i \phi_{12}} &J_{13} B_{13} e^{-i \phi_{13}} \\
	J_{12} B_{12} e^{i \phi_{12}} & (\nu_{2} - \omega_{d}) - i \frac{\kappa}{2} & J_{23} B_{23} e^{-i \phi_{23}} \\
	J_{13} B_{13} e^{i \phi_{13}} & J_{23} B_{23} e^{i \phi_{23}} & (\nu_{3} - \omega_{d}) - i \frac{\kappa}{2}
	\end{array}\right) \, .
\end{align}
The equations of motion for the phases $\phi_{ij}$ are unchanged. In Fig.~\ref{fig:OKdiss} we show the optical amplitude
$\left| \langle a_{j} \rangle \right|^{2} = \left| A_{j} \right|^{2}$ as a function of the flux $\Phi$ and the drive frequency $\omega_{d}$.
Figure~\ref{fig:OKdiss} almost resembles the eigenfrequencies $\Omega_{k}$ in Fig.~\ref{fig:circ} c) which we obtained from
diagonalizing the Hamiltonian.
%

\section{The Three-Site Model: Diagonalization}\label{appA}

Here, we give some further details on the three-site model given by Eq.~(\ref{eqn:eq1}) which in its explicit form reads
\begin{align}\label{eqn:S1}
	H &=	\sum_{j \in \{1,2,3\}} \nu_{j}^{\pd} a_{j}^{\dag} a_{j}^{\pd} + \sum_{l \in \{12,23,13\}} \omega_{l}^{\pd} b_{l}^{\dag} b_{l}^{\pd}
		+ J_{12}  \left( b_{12}^{\pd} a_{2}^{\dag} a_{1}^{\pd} + \Hc \right)
		+ J_{23}  \left( b_{23}^{\pd} a_{3}^{\dag} a_{2}^{\pd} + \Hc \right)
		+ J_{13}  \left( b_{13}^{\dag} a_{1}^{\dag} a_{3}^{\pd} + \Hc \right) \, .
\end{align}
The equations of motion are obtained straight forwardly by using Heisenberg's equation of motion. The mechanical phases
$\phi_{ij}$ evolve according to
\begin{align}
	\dot{\phi}_{12}	&= -\omega_{12} - \frac{J_{12} }{B_{12}} A_{1} A_{2} \cos(\phi_{12}+\theta_{12}) \, , \\
	\dot{\phi}_{23}	&= -\omega_{23} - \frac{J_{23} }{B_{23}} A_{2} A_{3} \cos(\phi_{23}+\theta_{23}) \, ,\\
	\dot{\phi}_{13}	&= -\omega_{13} - \frac{J_{13} }{B_{13}} A_{1} A_{3} \cos(\phi_{13}+\theta_{13}) \, .
\end{align}
The optical amplitudes $A_{j}$ obey the following equations of motion
\begin{align}
	\dot{A}_{1} =  &-J_{12}  B_{12} \sin(\phi_{12} + \theta_{12}) A_{2} - J_{13}  B_{13} \sin(\phi_{13} + \theta_{13}) A_{3} \, , \\
	\dot{A}_{2} =  &+J_{12}  B_{12} \sin(\phi_{12} + \theta_{12}) A_{1} - J_{23}  B_{23} \sin(\phi_{23} + \theta_{23}) A_{3} \, , \\
	\dot{A}_{3} =  &+J_{13}  B_{13} \sin(\phi_{13} + \theta_{13}) A_{1} + J_{23}  B_{23} \sin(\phi_{23} + \theta_{23}) A_{2} \, ,
\end{align}
and the optical phases $\theta_{j}$
\begin{align}
	\dot{\theta}_{1}	= - \nu_{1} &- J_{12}  B_{12} \frac{A_{2}}{A_{1}} \cos(\phi_{12} + \theta_{12}) - J_{13}  B_{13} \frac{A_{3}}{A_{1}} \cos(\phi_{13} + \theta_{13}) \, , \\
	\dot{\theta}_{2}	= - \nu_{2} &- J_{12}  B_{12} \frac{A_{1}}{A_{2}} \cos(\phi_{12} + \theta_{12}) - J_{23}  B_{23} \frac{A_{3}}{A_{2}} \cos(\phi_{23} + \theta_{23}) \, , \\
	\dot{\theta}_{3}	= - \nu_{3} &- J_{23}  B_{23} \frac{A_{2}}{A_{3}} \cos(\phi_{23} + \theta_{23}) - J_{13}  B_{13} \frac{A_{1}}{A_{3}} \cos(\phi_{13} + \theta_{13}) \, .
\end{align}

\begin{figure}[ht]
\begin{center}
	\includegraphics[width=0.6\columnwidth]{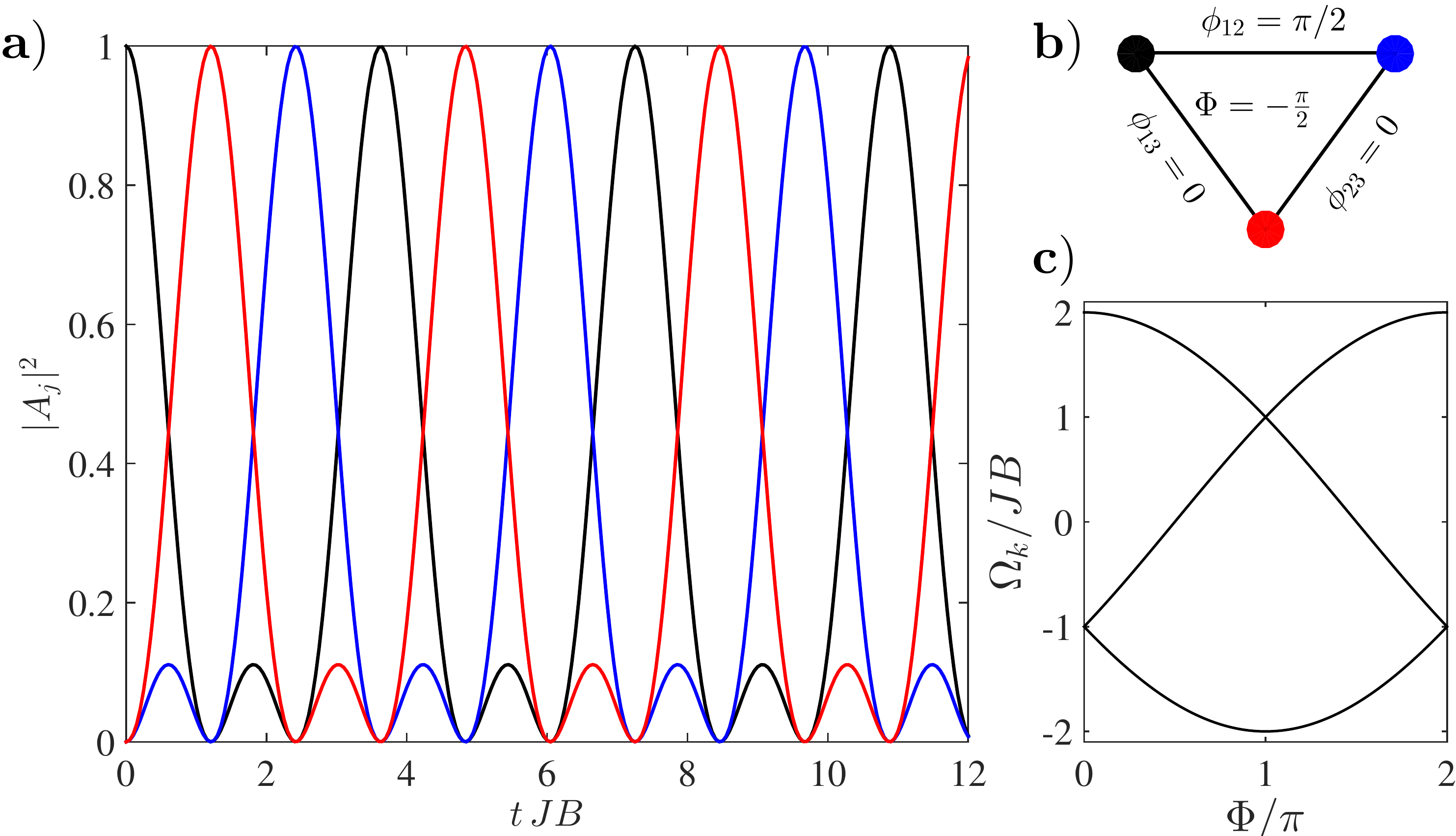}
	\caption{
	a) Optical amplitudes $|A_{j}|^{2}$. The initial state has one photon on site $1$. For $\Phi = -\pi/2$,
	the initial excitation moves counterclockwise around the triangular plaquette shown in b). For $\Phi = \pi/2$,
	the initial excitation would move clockwise around the triangle.
	c) Dispersion relation $\Omega_{k}$ as a function of the phase $\Phi$.
	}
	\label{fig:circ}
\end{center}
\end{figure}

As mentioned in the main text, for $A/B \ll 1$, i.e., in the case of a static flux $\Phi$, and for $\Phi = \pm \pi/2$
a circulator behavior is expected, cf.~Ref.~\onlinecite{Koch2010}. A solution to this system of coupled first order differential equations with
initially one photon on site $1$ and a phase $\Phi = - \pi/2$  is shown in Fig.~\ref{fig:circ} a) and b), where the circulator
behavior is clearly visible, i.e., the photon is moving counterclockwise around the triangular plaquette.

The Hamiltonian of the three-site model can be diagonalized best after going into a rotating frame by applying the
transformation
\begin{align}
	U = e^{i t \sum_{j} \nu_{j}^{\pd} a_{j}^{\dag} a_{j}^{\pd}} e^{i t \sum_{l} \omega_{l}^{\pd} b_{l}^{\dag} b_{l}^{\pd} } \, ,
\end{align}
to Eq.~(\ref{eqn:S1}) which leads to the Hamiltonian
\begin{align}
	H &=	J_{12}  B_{12} \left( e^{-i \phi_{12}} a_{1}^{\dag} a_{2}^{\pd} + \Hc \right) 
		+ J_{23}  B_{23} \left( e^{-i \phi_{23}} a_{2}^{\dag} a_{3}^{\pd} + \Hc \right) 
		+ J_{13}  B_{13} \left( e^{i \phi_{13}} a_{3}^{\dag} a_{1}^{\pd} + \Hc \right) \, .
\end{align}
Since we are here interested in the eigenvalues, we can perform a gauge transform $a_{j} \to a_{j} e^{i\chi_{j}}$ and make the
following gauge choice
\begin{align}\label{eqn:gif}
	- \phi_{12} + \chi_{2} - \chi_{1} &=  \tilde{\Phi} \, , \\
	- \phi_{23} + \chi_{3} - \chi_{2} &= \tilde{\Phi} \, , \\
	\phi_{13} + \chi_{1} - \chi_{3}   &= \tilde{\Phi} \, , 
\end{align}
which we can write as $\phi_{13} - \phi_{12} - \phi_{23} = 3 \tilde{\Phi} = \Phi$.
The Hamiltonian can then be written as
\begin{align}
	H = \sum_{j=1}^{3} J_{j} B_{j} e^{i \tilde{\Phi}} a_{j}^{\dag} a_{j+1}^{\pd} + \Hc \, ,
\end{align}
where we assumed periodic boundary conditions, i.e., $a_{4} = a_{1}$ and for convenience we introduce $J_{12,23,13} \to J_{1,2,3}$
and similarly for $B_{12,23,13}$. By introducing normal modes
\begin{align}
	a_{j}^{\dag} = \frac{1}{\sqrt{3}} \sum_{k=0}^{2} e^{-2 \pi i k j / 3} A_{k}^{\dag} \, , \\
	A_{k}^{\dag} = \frac{1}{\sqrt{3}} \sum_{j=1}^{3} e^{2 \pi i k j / 3} a_{j}^{\dag} \, ,
\end{align}
and furthermore assuming equal tunneling amplitude $J$ and limit-cycle amplitude $B$, the Hamiltonian can be diagonalized
\begin{align}
	H = \sum_{k=0}^{2} \Omega_{k}^{\pd} A_{k}^{\dag} A_{k}^{\pd} \, ,
\end{align}
where $\Omega_{k} = 2 J B \cos \left( \Phi/3 + 2 \pi k/3  \right)$. As already pointed out in Ref.~\onlinecite{Koch2010}, for $\Phi = \pm\pi/2$
this setup shows the behaviour of a photon circulator.

\section{Spectrum of a Periodic Trajectory Versus a Chaotic Trajectory}\label{appC}
%
\begin{figure}[ht]
\begin{center}
	\includegraphics[width=0.8\columnwidth]{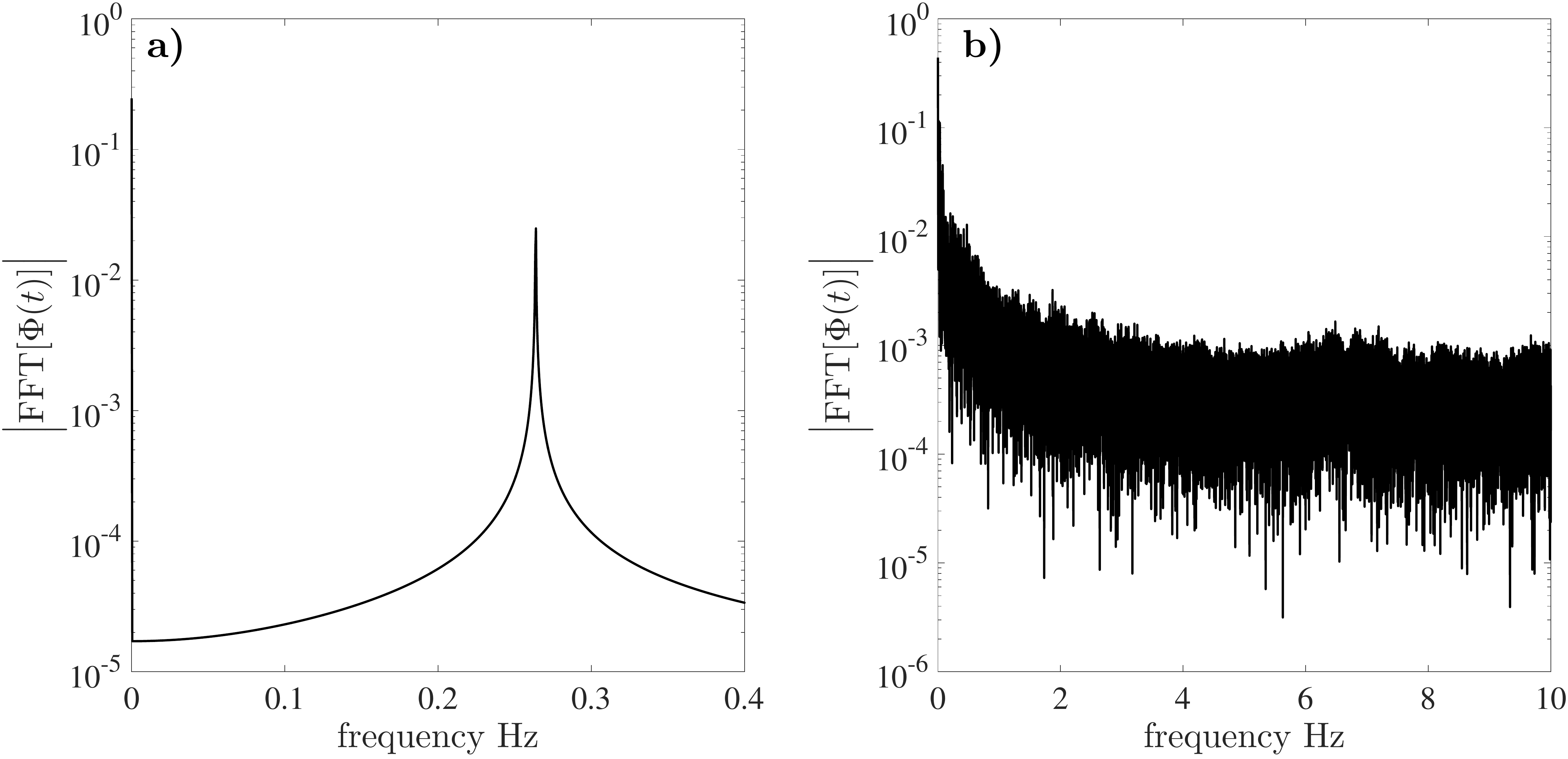}
	\caption{
	Fourier transform of the trajectories shown in Fig.~\ref{fig:fig22} a) and Fig.~\ref{fig:fig22} b). In a) the spectrum shows
	a clear peak indicating a periodic trajectory, cf. Fig.~\ref{fig:fig22} a). In b) however the spectrum is flat which
	indicates a chaotic trajectory, cf. Fig.~\ref{fig:fig22} b).
	}
	\label{fig:FFT}
\end{center}
\end{figure}

Here, we give additional information on how we made the distinction between the periodic and chaotic behavior of the
dynamical gauge field in the case of the three-site model. In the main text, we chose to present the distinction, periodic
versus chaotic, by plotting the bifurcation diagram shown in Fig.~\ref{fig:fig22} c). In addition to this, we also investigated the Fourier
transform of the time series $\Phi(t)$, the spectrum. For a periodic time series there is a pronounced peak in the
spectrum at the frequency corresponding to the period of the time series. In the case of a chaotic time series the
spectrum does not show a preferred frequency and is almost flat.
In Fig.~\ref{fig:FFT} a) and Fig.~\ref{fig:FFT} b), we show the Fourier transform of the time series corresponding to
the phase space trajectories shown in Fig.~\ref{fig:fig22} a) and Fig.~\ref{fig:fig22} b), respectively. In the case of a periodic
trajectory, the spectrum clearly shows a peak, where as for a chaotic trajectory the spectrum is flat.
To summarize, in addition to the bifurcation diagram presented in the main text, we also checked the Fourier transform
of the time series to consistently distinguish periodic from chaotic trajectories.

\section{The Triangular Lattice: Equations of Motion}\label{appD}

On a two-dimensional triangular lattice we denote dynamical variables on lattice site $n,m$ with a subscripts $(n,m)$ and
dynamical variables on the directed links from site $n,m$ to site $k,l$ by a subscript $(n,m)(k,l)$. The equations of motion
for the mechanical phases $\phi_{(n,m)(n,m+1)}$ on the lattice links are
\begin{align}
	 &\dot{\phi}_{(n,m)(n,m+1)} = -\omega_{(n,m)(n,m+1)} - \frac{J_{(n,m)(n,m+1)}}{B_{(n,m)(n,m+1)}} A_{n,m} A_{n,m+1} \cos(\phi_{(n,m)(n,m+1)}+\theta_{n,m}-\theta_{n,m+1}) \, , \nn \\
	 &\dot{\phi}_{(n,m+1)(n+1,m)}	= -\omega_{(n,m+1)(n+1,m)}- \frac{J_{(n,m+1)(n+1,m)}}{B_{(n,m+1)(n+1,m)}} A_{n,m+1} A_{n+1,m} \cos(\phi_{(n,m+1)(n+1,m)}+\theta_{n,m+1}-\theta_{n+1,m}) \, , \nn \\
	 &\dot{\phi}_{(n,m)(n+1,m)}		= -\omega_{(n,m)(n+1,m)} - \frac{J_{(n,m)(n+1,m)}}{B_{(n,m)(n+1,m)}} A_{n,m} A_{n+1,m} \cos(\phi_{(n,m)(n+1,m)}+\theta_{n,m}-\theta_{n+1,m}) \, . \nn
\end{align}
The optical amplitudes $A_{n,m}$ obey the equations of motion
\begin{align}
	\dot{A}_{n,m} = &-J_{(n,m)(n,m+1)} B_{(n,m)(n,m+1)} \sin(\phi_{(n,m)(n,m+1)} + \theta_{n,m} - \theta_{n,m+1}) A_{n,m+1}  \\
				&- J_{(n,m)(n+1,m)} B_{(n,m)(n+1,m)} \sin(\phi_{(n,m)(n+1,m)} + \theta_{n,m} - \theta_{n+1,m}) A_{n+1,m} \nn \\
				&+J_{(n,m-1)(n,m)} B_{(n,m-1)(n,m)} \sin(\phi_{(n,m-1)(n,m)} + \theta_{n,m-1} - \theta_{n,m}) A_{n,m-1} \nn \\
				&- J_{(n,m)(n+1,m-1)} B_{(n,m)(n+1,m-1)} \sin(\phi_{(n,m)(n+1,m-1)} + \theta_{n,m} - \theta_{n+1,m-1}) A_{n+1,m-1} \nn \\
				&+J_{(n-1,m)(n,m)} B_{(n-1,m)(n,m)} \sin(\phi_{(n-1,m)(n,m)} + \theta_{n-1,m} - \theta_{n,m}) A_{n-1,m} \nn \\
				&+ J_{(n-1,m+1)(n,m)} B_{(n-1,m+1)(n,m)} \sin(\phi_{(n-1,m+1)(n,m)} + \theta_{n-1,m+1} - \theta_{n,m}) A_{n-1,m+1} \, , \nn
\end{align}
and the optical phases $\theta_{n,m}$ on a lattice site evolve according to
\begin{align}
	\dot{\theta}_{n,m} = 	&- \nu_{n,m}  \\
					&- J_{(n,m)(n,m+1)} B_{(n,m)(n,m+1)} \frac{A_{n,m+1}}{A_{n,m}} \cos(\phi_{(n,m)(n,m+1)} + \theta_{n,m} - \theta_{n,m+1})  \nn \\
					&- J_{(n,m)(n+1,m)} B_{(n,m)(n+1,m)} \frac{A_{n+1,m}}{A_{n,m}} \cos(\phi_{(n,m)(n+1,m)} + \theta_{n,m} - \theta_{n+1,m}) \nn \\
					& - J_{(n,m-1)(n,m)} B_{(n,m-1)(n,m)} \frac{A_{n,m-1}}{A_{n,m}} \cos(\phi_{(n,m-1)(n,m)} + \theta_{n,m-1} - \theta_{n,m}) \nn \\
					&- J_{(n,m)(n+1,m-1)} B_{(n,m)(n+1,m-1)} \frac{A_{n+1,m-1}}{A_{n,m}} \cos(\phi_{(n,m)(n+1,m-1)} + \theta_{n,m} - \theta_{n+1,m-1}) \nn \\
					& - J_{(n,m)(n+1,m-1)} B_{(n,m)(n+1,m-1)} \frac{A_{n-1,m}}{A_{n,m}} \cos(\phi_{(n,m)(n+1,m-1)} + \theta_{n-1,m} - \theta_{n,m}) \nn \\
					&- J_{(n-1,m+1)(n,m)} B_{(n-1,m+1)(n,m)} \frac{A_{n-1,m+1}}{A_{n,m}} \cos(\phi_{(n-1,m+1)(n,m)} + \theta_{n-1,m+1} - \theta_{n,m}) \,  \nn.
\end{align}
A driving term on a particular site and dissipation of the photons on each site can be added straightforwardly, as also done
for the three-site model.

\section{The Triangular Lattice: Distribution of Optical and Mechanical Frequencies}\label{appE}

\begin{figure}[ht!]
\begin{center}
	\includegraphics[width=0.35\columnwidth]{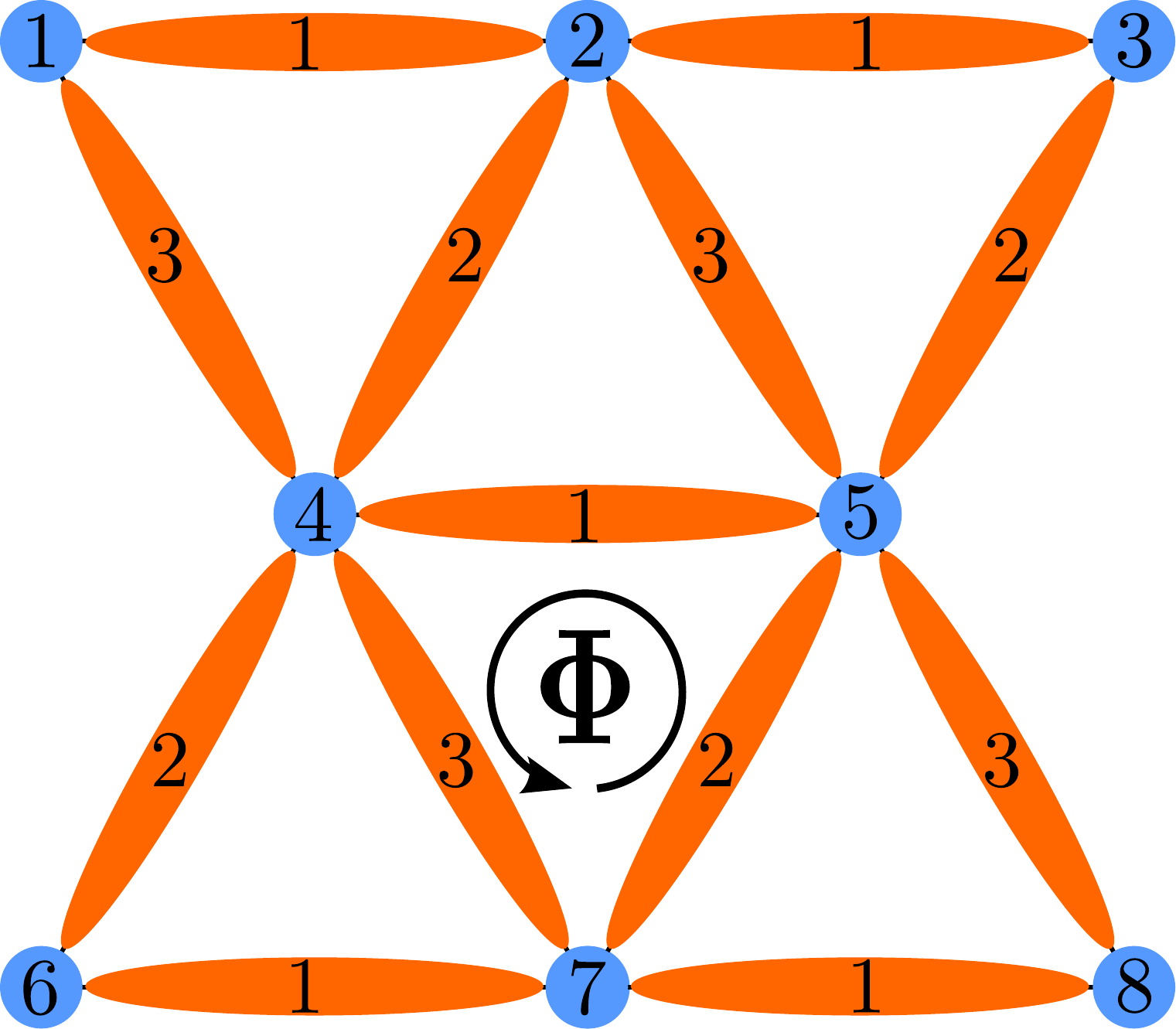}
	\caption{
	Possible distribution of optical (blue dots) and mechanical (orange ellipses) frequencies on a $3\times3$ triangular lattice as
	an example.
	}
	\label{fig:FE}
\end{center}
\end{figure}

As mentioned in the main text, the links are directed: A photon tunneling from a site with a low (high) on-site frequency to
a site with a high (low) on-site frequency absorbs (emits) a phonon. These processes have to be resonant which could be
achieved by distributing the photon and phonon frequencies in the right way. For instance, the frequencies can be arranged
as shown in Fig.~\ref{fig:FE}.
%
%




\end{document}